\documentclass[onecolumn]{IEEEtran}
\usepackage{booktabs}
\usepackage{hhline}
\usepackage{cite}
\usepackage[T1]{fontenc}
\usepackage{bbm}
\usepackage{amssymb}
\usepackage{amsfonts}
\usepackage{epsfig}
\usepackage{calc}
\usepackage{color}
\usepackage[all]{xy}
\usepackage{bm}
\usepackage{algpseudocode}
\usepackage{enumerate}
\usepackage{pdfsync}
\usepackage{framed}
\usepackage{caption}
\usepackage{amsthm}
\usepackage{blkarray}
\usepackage{subcaption}
\usepackage{enumitem}
\usepackage[linesnumbered,ruled,vlined]{algorithm2e}

\usepackage{mathrsfs}
\usepackage{caption}
\usepackage{float}
\newtheorem{defn}{Definition}[section]

\newtheorem{thm}{Theorem}%[section]
\newtheorem{cor}[thm]{Corollary}
\newtheorem{prop}{Proposition}

\newtheorem{lem}[thm]{Lemma}
\newtheorem{conj}[thm]{Conjecture}
\newtheorem{constr}[thm]{Construction}
\usepackage[normalem]{ulem}

\newtheorem{remark}{Remark}%[section]

\newcounter{definition}[section]

\newcommand{\bit}{\begin{itemize}}
	\newcommand{\eit}{\end{itemize}}
\newcommand{\bcor}{\begin{cor}}
	\newcommand{\ecor}{\end{cor}}
\newcommand{\beq}{\begin{equation}}
	\newcommand{\eeq}{\end{equation}}
\newcommand{\beqn}{\begin{equation*}}
	\newcommand{\eeqn}{\end{equation*}}
\newcommand{\bea}{\begin{eqnarray}}
	\newcommand{\eea}{\end{eqnarray}}
\newcommand{\bean}{\begin{eqnarray*}}
	\newcommand{\eean}{\end{eqnarray*}}
\newcommand{\ben}{\begin{enumerate}}
	\newcommand{\een}{\end{enumerate}}
\newcommand{\bdefn}{\begin{defn}}
	\newcommand{\edefn}{\end{defn}}
\newcommand{\bnote}{\begin{remark}}
	\newcommand{\enote}{\end{remark}}
\newcommand{\bprop}{\begin{prop}}
	\newcommand{\eprop}{\end{prop}}
\newcommand{\blem}{\begin{lem}}
	\newcommand{\elem}{\end{lem}}
\newcommand{\bthm}{\begin{thm}}
	\newcommand{\ethm}{\end{thm}}
\newcommand{\bconj}{\begin{conj}}
	\newcommand{\econj}{\end{conj}}
\newcommand{\bconstr}{\begin{constr}}
	\newcommand{\econstr}{\end{constr}}
\newcommand{\bpf}{\begin{proof}}
	\newcommand{\epf}{\end{proof}}

\usepackage[utf8]{inputenc} 
\usepackage[T1]{fontenc}
\usepackage{url}              % provides \url{...}
\usepackage{cite}             % improves presentation of citations

\usepackage[cmex10]{amsmath}  % Use the [cmex10] option to ensure complicance
\usepackage{mleftright}       % fix to wrong spacing of \left-,
\mleftright                   % \middle- \right-commands 

\usepackage{graphicx}         % provides \includegraphics{...} to
\usepackage{booktabs}         % fixes poor spacing in tables and
\begin{document}

\title{Hierarchical Coded Gradient Aggregation Based on Layered MDS Codes} 

%%%%%%
\author{ 
	\IEEEauthorblockN{	
		M. Nikhil Krishnan\IEEEauthorrefmark{1}, 
		Anoop Thomas\IEEEauthorrefmark{2},
		Birenjith Sasidharan\IEEEauthorrefmark{3}\\}
	\IEEEauthorblockA{\IEEEauthorrefmark{1}%
		Department of Data Science, IIT Palakkad\\}
	\IEEEauthorblockA{\IEEEauthorrefmark{2}%
		School of Electrical Sciences, IIT Bhubaneswar\\}
	\IEEEauthorblockA{\IEEEauthorrefmark{3}	Department of Electrical and Computer Systems Engineering, Monash University}
	\IEEEauthorblockA{email: 
		nikhilkrishnan.m@gmail.com, anoopthomas@iitbbs.ac.in, birenjith@gmail.com}
	\thanks{This work was supported by faculty seed grants from the Indian Institute of Technology Palakkad, Kerala, India (to M. N. Krishnan) and the Indian Institute of Technology Bhubaneswar, Odisha, India (to A. Thomas). M. N. Krishnan also acknowledges the support from the DST-INSPIRE fellowship. B. Sasidharan is on leave of absence from the Government Engineering College, Barton Hill (APJAK Technological University), Kerala, India.}
}

\IEEEoverridecommandlockouts
\maketitle

%%%%%
%% Abstract: 
%% If your paper is eligible for the student paper award, please add
%% the comment "THIS PAPER IS ELIGIBLE FOR THE STUDENT PAPER
%% AWARD." as a first line in the abstract. 
%% For the final version of the accepted paper, please do not forget
%% to remove this comment!
%%
\begin{abstract}
	The growing privacy concerns and the communication costs associated with transmitting raw data have resulted in techniques like federated learning, where the machine learning models are trained at the edge nodes, and the parameter updates are shared with a central server. Because communications from the edge nodes are often unreliable, a hierarchical setup involving intermediate helper nodes is considered. The communication links between the edges and the helper nodes are error-prone and are modeled as straggling/failing links. To overcome the issue of link failures, coding techniques are proposed. The edge nodes communicate encoded versions of the model updates to the helper nodes, which pass them on to the master after suitable aggregation. The primary work in this area uses repetition codes and Maximum Distance Separable (MDS) codes at the edge nodes to arrive at the Aligned Repetition Coding (ARC) and Aligned MDS Coding (AMC) schemes, respectively. We propose using vector codes, specifically a family of layered MDS codes parameterized by a variable $\nu$, at the edge nodes.  For the proposed family of codes, suitable aggregation strategies at the helper nodes are also developed. At the extreme values of $\nu$, our scheme matches the communication costs incurred by the ARC and AMC schemes, resulting in a graceful transition between these schemes.

\end{abstract}

\section{Introduction\label{sec:intro}}

Modern machine learning applications operate on a large amount of data that is generated at the edge/client nodes. Typically, the learning models are trained on a central server, and the data collected at the edge nodes is transmitted to the server. The above method has two serious shortcomings. The amount of data is huge and transmitting such data involves a lot of communication costs for the edge nodes, which are often constrained by power and bandwidth. The second shortcoming is with respect to data privacy and there is a growing concern over sharing raw data with the central server. These factors have led to the development of the Federated Learning (FL) framework\cite{McMRHA2017c,LiSTS20, ReMHP2020,ZhWZZC2020}, wherein models are trained directly on the edge devices and only model updates are transmitted to the server. Since only the updates are transmitted, the technique significantly reduces transmission costs and mitigates privacy concerns. 

Gradient descent is a widely used algorithm for training machine learning models. Synchronous gradient descent over centralized data using multiple worker nodes is well-studied, and algorithms to improve the computation time have been proposed in the literature (for instance, see \cite{TaLDK2017,YeA2018,LiMMYSA2018,RePPA2019,KHKJSIT21}). The current paper considers synchronous gradient descent over a decentralized data set in the FL setting. The data sets are available only locally at the edge nodes, which compute gradients over their respective data sets. The computed gradients are then shared with a central server for a global model update. The edge nodes are often power-constrained and may only be available intermittently. As a result, the communications between edge nodes and the central server are unreliable. Hence, to assist the edge nodes, hierarchical systems are considered in \cite{PrRPA2020, LZSL20}. In this paper, we adopt the hierarchical coded gradient aggregation strategy proposed in \cite{PrRPA2020}, where intermediate helper nodes support the edge nodes in communicating the computed gradient vectors to the central server. Each edge node computes the gradient over its data set and transmits coded fragments of the gradient to all the helper nodes. These coded fragments are obtained using a suitable linear code, referred to as the client code. The transmissions from edge nodes to the helper nodes are assumed to be unreliable and this scenario is modeled by introducing straggling links. Typically, the helper nodes have more computing and communication resources and hence the transmissions between the helper nodes and the central server are assumed to be error-free. Moreover, to reduce communication costs, the helper nodes communicate the received gradient vectors to the central server possibly after an aggregation step. 

The paper \cite{PrRPA2020} proposes two hierarchical coded gradient aggregation schemes that use two different client codes; the Aligned Repetition Coding (ARC) scheme, based on repetition codes and the Aligned MDS Coding (AMC) scheme, based on Maximum Distance Separable (MDS) codes. The ARC scheme allows reducing the helper-to-master communication cost while performing poorly with respect to the edge-to-helper communication cost. The AMC scheme, on the other hand, aims at reducing the edge-to-helper communication cost to the lowest possible value. In a scheme proposed in \cite{BTJSACCGA} that makes use of pyramid codes\cite{HuaCL2007}, it is possible to achieve a reduction in helper-to-master communication cost at the expense of edge-to-helper communication cost, by tuning a certain parameter of the scheme.
 	
We observe that previous works have solely considered scalar codes as client codes. Vector codes, a more general coding framework that permits to have vectors as code symbols, have been proved useful in other contexts such as design of low-complexity MDS codes \cite{BlaBV1996}, regenerating codes \cite{DimGWWR2010} etc. In the present work, we propose to use vector codes as client codes and they turn out to do well here as well. In the remainder of this section, we describe the system model, formalize the problem setting and present a summary of our main results.

\subsection{System Model\label{sec:sys}}

	We begin with some notation. Let $\mathbb{Z}$ denote the set of all integers. For $i,j \in\mathbb{Z}$, we have $[i:j]\triangleq \{x\in\mathbb{Z}\mid i\leq x\leq j\}$. As a special case, we use $[j]$ to indicate the set $[1:j]$. Given a matrix $A$, we use $A(i,j)$ to denote the element which lies in row $i$ and column $j$.

	There is a central server or a master $M$ aiming to compute an optimal vector $w^*\in\mathbb{R}^p$ (i.e., there are $p$ parameters for the learning model) based on data samples available locally at $n_e$ edge (client) nodes. Let $D$ denote the collection of all data samples distributed across edge nodes. The function $\sum_{x \in D} \ell(w,x)$ is to be minimized to compute the optimal $w^*$, where $\ell(.)$ is an underlying loss function. The minimization can be done by carrying out a gradient descent algorithm in which every iteration $t$ involves an update rule of $w^{t+1} = w^t - \mu \sum_{x \in D} \nabla\ell(w^t,x)$, where $\mu$ is the learning rate. Let $D = D_1 \cup D_2 \cup \cdots \cup D_{n_e}$ be a partition of data samples, where $D_i$ denotes the subset of data samples available at the $i$-th edge node $E_i$. We can then write the (global) gradient as $\underline{g} := \sum_{x \in D} \nabla\ell(w^t,x) = \sum_{i=1}^{n_e}\sum_{x \in D_i} \nabla\ell(w^t,x) = \sum_{i=1}^{n_e} \underline{g}_i$, and we define $\underline{g}_i \in \mathbb{R}^p$ as the local gradient available at the $i$-th edge node. 

%Each edge node transmits $\underline{g}_i$ at start of every iteration and the master sums up the local gradients to compute the global gradient. 

\begin{figure}[!]
	\begin{center}
		\includegraphics[scale=0.5]{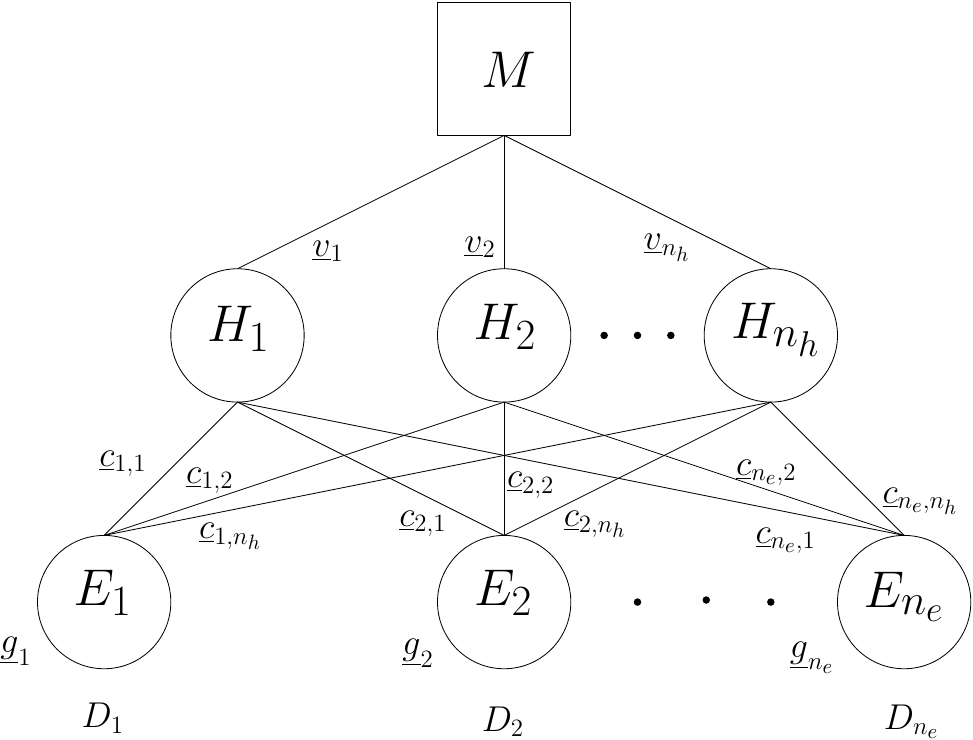}
		\caption{The hierarchical coded gradient aggregation setting with $n_e$ edge nodes and $n_h$ helper nodes.}
		\label{fig:setup}
	\end{center}
%	\vspace{-0.2in}
\end{figure}

In the hierarchical model, the $n_e$ edge nodes $E_1,E_2,\ldots,E_{n_e}$ are connected to the master via $n_h$ helper nodes $H_1, H_2, \ldots, H_{n_h}$. Every edge is connected to every helper, and every helper in turn to the master. This is illustrated in Fig. \ref{fig:setup}. A helper is not just a router, but can as well process the messages it receives, and pass them to the master. For this reason, they are also called aggregators. The links from helpers to the master are assumed to be perfectly reliable, whereas those from edges to helpers are not. We consider the following adversarial channel model between edge nodes and helpers. Among the $[n_h]$ links available to each edge node (one link to each helper node), up to $s \in [n_h-1]$ can straggle or fail. Each helper node communicates to the master the observed erasure (straggler) pattern, represented by a binary vector of length $n_e$. The master constructs an erasure matrix ${\cal E}$, an $n_e \times n_h$ binary matrix, where ${\cal E}(i,j)=1$ if and only if the link between edge  $E_i$ and helper $H_j$ has failed. The erasure matrix ${\cal E}$ is passed to all the helpers. Thus, all helpers and the master have full knowledge of the erasure matrix.  

While the gradient vectors are indeed real vectors, they are often quantized and represented by a vector over a finite field. The effect of quantization on the convergence of training and other related issues forms a separate line of research \cite{AlGLTV2017,AcDFS2019,ShCEPC2020}, which is not of interest in this paper. We henceforth assume that all the mathematical operations are over finite fields. Specifically, we set $\underline{g},\underline{g}_i \in \mathbb{F}^p$, where $\mathbb{F}$ is the finite field over which the client code is defined.

Let alphabet $\mathbb{A} \triangleq \mathbb{F}^d$, where $d$ is a parameter dependent on the client code being used. An edge $E_i$ generates $\underline{c}_{i,j} \in \mathbb{A}^{b_{i,j}}$ as a function of $\underline{g}_i$ and transmits $\underline{c}_{i,j}$ to $H_j$, for every $j\in[n_h]$. We assume that every edge node uses the same function in this process. We shall restrict to linear functions over $\mathbb{F}$ and also fix $b_{ij}=b$ for every $i,j$. We describe the computation of $\underline{c}_{i,j}$ in terms of a linear code ${\cal C}$ over $\mathbb{F}$. The code ${\cal C}$ is referred to as the {\em client code}. 

%An edge $E_i$ splits the local gradient $\underline{g}_i$ into $k$ chunks $\underline{g}_{i,j} \in \mathbb{R}^{(p/k)}, j \in [k]$. These chunks are encoded using a linear block code (termed as {\em client code}) of length $n_h$ and dimension $k$ with generator matrix $G$ to obtain $\underline{c}_i  =  [\underline{g}_{i,1} \ \underline{g}_{i,2} \ \cdots \underline{g}_{i,k}] \cdot G ,  \forall i \in [n_e], $ where $\underline{c}_i = [\underline{c}_{i,1} \ \underline{c}_{i,2} \ \cdots \underline{c}_{i,n_h}]  \in \mathbb{R}^{(p/k) \times n_h}$. Each edge node $E_i, i \in [n_e]$ transmits the coded chunk $\underline{c}_{i,j}$ to the helper node $H_j$ for all $j \in [n_h]$. 

%\label{sec:setting}
%Let $n_e,n_h$ denote the number of edge nodes and helper nodes, respectively. Let $\underline{g}_i\in\mathbb{R}^{p\times 1}$  denote the ``local gradient'' of size $p$ computed by edge node $i$, where $i\in[n_e]$. Each edge node $i$ sends $b_{i,j}$ symbols (i.e., real numbers) to helper node $j$, $j\in[n_h]$. These $b_{i,j}$ symbols are obtained as a function of $\underline{g}_i$. For simplicity, we assume $b_{i,j}\triangleq b$, which is a constant regardless of the choice of $i$ and $j$. We consider an adversarial channel model wherein up to $s$ transmissions (out of $n_h$) made by an edge node to a helper node can possibly fail. Excluding the trivial scenarios of $s=0$ (no transmission can fail) and $s=n_h$ (all transmissions can fail), we have $s\in[n_h-1]$.

Let $\Omega(s)$ denote the set of all erasure matrices with $s$ straggling links per edge node. Assume that an erasure matrix $\epsilon \in \Omega(s)$ has occurred and is known to all the helpers. Given an $\epsilon \in \Omega(s)$, every helper node $H_j$ invokes a function ${\cal A}_j$ that maps $\{\underline{c}_{i,j}\mid i\in[n_e],\epsilon(i,j)=0\}$ to a vector $\underline{v}_j \in \mathbb{A}^{m_{j}(\epsilon)}$ and transmits $\underline{v}_j$ to the master. The collection of functions ${\cal A}= ({\cal A}_1,{\cal A}_2, \ldots, {\cal A}_{n_h})$ is referred to as an {\em aggregation strategy}. We emphasize that both $\mathcal{A}_j$ and $m_{j}(\epsilon)$ depend on the realization $\epsilon$ of the erasure matrix. The messages $\{\underline{v}_j\}$ to the master are assumed to be transmitted over error-free links. The master makes use of $\{\underline{v}_j\mid j\in[n_h]\}$ to compute $\underline{g} =  \sum_{i=1}^{n_e} \underline{g}_{i}$. The $2$-tuple $({\cal C},{\cal A})$ is referred to as a {\it scheme} for coded gradient aggregation.

\subsection{Communication Costs}

Communication costs are quantified by the number of transmitted symbols from the field $\mathbb{F}$, normalized by the number of parameters (gradient vector size) $p$. An edge node transmits $b$ symbols over the alphabet $\mathbb{A}\triangleq \mathbb{F}^d$ to every helper. Therefore, we define the communication cost from edge to helpers as:
\bean
C_{\text{EH}} & = & \frac{n_hdb}{p}.
\eean
Assume that helpers employ an aggregation strategy ${\cal A}$. Since $m_j(\epsilon)$ is a function of the erasure matrix $\epsilon$, the communication cost from helpers to the master in turn, is a function of $\epsilon$ and is defined as:
\bean
C_{\text{HM}}(\epsilon) & = & \sum_{j=1}^{n_h} \frac{dm_j(\epsilon)}{p}.
\eean 
If we impose a uniform probability distribution over $\Omega(s)$, 
the average communication cost from helpers to the master can naturally be defined as:
\bean
C_{\text{HM},\text{av}} = \ \mathbb{E}[C_{\text{HM}}(\epsilon)] \ = \ \frac{1}{|\Omega(s)|} \sum_{\epsilon \in \Omega(s)} \sum_{j=1}^{n_h} \frac{dm_j(\epsilon)}{p}.
\eean 
Alternatively, one may pursue a worst-case approach and define the communication cost from helpers to the master as:
\bean 
C_{\text{HM}} = \max_{\epsilon \in \Omega(s)} \sum_{j=1}^{n_h} \frac{dm_j(\epsilon)}{p}.
\eean 
In this paper, we shall adopt the worst-case approach in our analysis of communication costs. Since $C_\text{HM}$ itself is a function of $\mathcal{C}$ and $\mathcal{A}$, we also observe that it may be meaningful to define the min-max cost: 
\bean
C_{\text{HM}}^* & = & \min_{{\left(\mathcal{C}, \cal A \right)}} C_{\text{HM}}.
\eean 
 However, we will not be pursuing this direction in the current paper.

%\subsection{Single-max and Multi-max Aggregation Strategies\label{sec:ags}}

\subsection{Our Contributions \label{sec:oc}}

\bit
\item We propose the use of vector codes (i.e., $b\geq 1$) as client codes for the hierarchical coded gradient aggregation framework. In particular, we explore the idea of layering numerous short-block-length MDS codes to create a vector code (for this reason, we will refer to the resultant client code as a layered code). We attribute this idea to the work \cite{TBAPTIT}, which used it in the context of distributed storage systems (DSSs). In the DSS literature, this idea is known \cite{SPKVSK16} to result in optimal operating points on a storage overhead vs. repair bandwidth trade-off curve. 
  
\item The proposed layered code is parameterized by $\nu\in[n_h-s]$. Our code is equivalent to that used in the ARC scheme at $\nu=1$, while at $\nu=n_h-s$, the code reduces to that used in the AMC scheme. 

\begin{figure}[!]
	\begin{center}
		\includegraphics[scale=0.45]{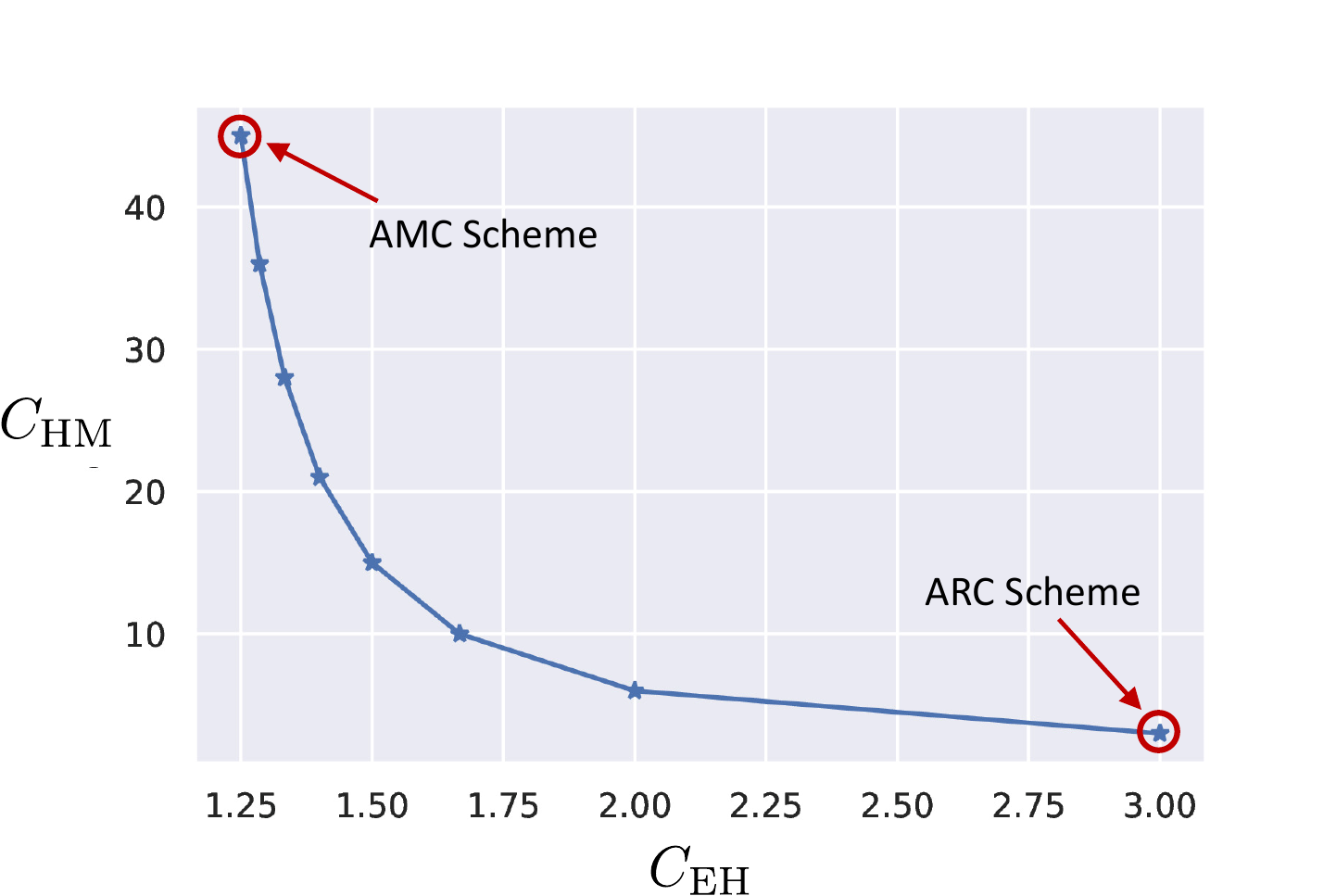}
		\caption{The tradeoff existing between $C_\text{EH}$ and $C_\text{HM}$. Here, $n_h=10$, $s=2$ and $n_e=50$. Our scheme has a certain parameter $\nu\in[8]$. As $\nu$ increases, the operating point moves from right to left in the curve (trading off $C_\text{HM}$ for reducing $C_\text{EH}$).}
		\label{fig:tradeoff}
	\end{center}
	%\vspace{-0.2in}
\end{figure}

\item We propose a novel aggregation strategy, which results in achieving intermediate operating points between the ARC and AMC schemes by trading off $C_\text{HM}$ for $C_\text{EH}$. For instance, see Fig.~\ref{fig:tradeoff}.
While a few intermediate points close to the AMC are provided in \cite{BTJSACCGA}, our proposed scheme achieves a graceful transition between the extreme points.

\eit

%\bit
%
%\item 
%\item parametrized scheme
%\item array --> vector code, layered code, scheme, aggregation.
%\item We achieve a graceful transition from amc to arc while trading off.....  refer to figure, corner points
%\eit
%
%We consider vector codes for client codes in this framework, i.e., the number of symbols over the alphabet $\mathbb{A}$ transmitted by each edge node is $b\geq 1$. This is unlike previous works which consider the case of $b=1$.
%
%Motivated by an earlier work in the distributed storage literature which uses layered MDS codes \cite{Tianetal}, we consider the layering of MDS-codes to generate our client vector code. The advantages are multi-fold. 
%
%\bit
%\item We achieve a trade-off between $C_\text{EH}$ and $C_{HM}$ (see Fig.~\ref{fig:tradeoff}). Specifically, our client code has a parameter $\nu\in[1:n_h-s]$. Our scheme reduces to the ARC scheme at $\nu=1$, whereas at $\nu=n-s$, it is equivalent to the AMC scheme.
%\item As we use multiple layers of short-block-length MDS codes in our coding scheme, it facilitates ease of implementation. 
%\eit

\section{The Layered Coding Scheme\label{sec:coding_scheme}}

The scheme referred to as the {\em layered coding scheme} consists of a client code ${\cal C}_\nu$ and an aggregation strategy ${\cal A}_\nu$. Unlike previously known schemes with scalar codes, we propose to use a vector code as the client code, i.e., with $b \geq 1$. %We will also elucidate the scheme with the help of an example.

\subsection{Description of ${\cal C}_\nu$\label{sec:scheme}}

%{\color{red} Need to highlight subpacketization}
Let the parameters of the system model $(p,n_e,n_h,s)$ be given. Fix the parameter $\nu \in [n_h-s]$ and set $L = {n_h \choose \nu+s}$ and $\lambda = L\nu$. Here, $L$ denote the number of layers (to be defined shortly). We choose the alphabet of the client code $\mathbb{A} = \mathbb{F}^d$ where $d = \frac{p}{\lambda}$.  The client code ${\cal C}_\nu$ is a vector code and its construction is described by providing the explicit mapping from a message $\underline{g}_i \in \mathbb{F}^p$ to a codeword array $\underline{c}_{i} \in \mathbb{A}^{b \times n_h}$. Here $b$ is the subpacketization level of the vector code and its value will be evident soon. The mapping from $\underline{g}_i$ to $\underline{c}_i$ is described in three steps:

\ben
\item 
The vector $\underline{g}_i$ is partitioned into $\lambda$ subvectors $\{\underline{g}^{(\ell)}_{i,j} \mid j \in[\nu], \ell\in[L]\}$ each belonging to $\mathbb{A}$. Let ${\bf G}$ denote a $\nu \times (\nu+s)$ generator matrix of a $[\nu+s,\nu]$-MDS code $C_{\text{MDS}}$ over $\mathbb{F}$. Then for every $\ell\in[L]$, we define:
\bean
[\underline{c}^{(\ell)}_{i,1} \ \underline{c}^{(\ell)}_{i,2} \ \cdots \  \underline{c}^{(\ell)}_{i,\nu+s}] = [\underline{g}^{(\ell)}_{i,1} \ \underline{g}^{(\ell)}_{i,2} \ \cdots \ \underline{g}^{(\ell)}_{i,\nu}] {\bf G} .
\eean 

\item In this step, we shall arrange the symbols $\{\underline{c}^{(\ell)}_{i,j} \mid j \in[\nu+s], \ell\in[L]\}$ in an $L \times n_h$ array with certain cells left blank. Each row of this array is referred to as a layer, and thus there are $L$ layers. A cell of the array is indexed by a $2$-tuple $(x,y)$, where $x\in[L]$ denotes the row number and $y\in[n_h]$ is the column number. Let $\mathcal{H} = \{\mathcal{H}_1,\mathcal{H}_2,\ldots,\mathcal{H}_{L}\}$ denote the set of all $(\nu+s)$-element subsets of $[n_h]$ arranged in a lexicographic order. Every $\mathcal{H}_{\ell} \in \mathcal{H}$ is denoted by $\mathcal{H}_{\ell} = \{\mathcal{H}_{\ell,1},\mathcal{H}_{\ell,2}, \cdots, \mathcal{H}_{\ell,\nu+s}\}$ such that  $\mathcal{H}_{\ell,1} < \mathcal{H}_{\ell,2} < \cdots < \mathcal{H}_{\ell,\nu+s}$. Then $\underline{c}^{(\ell)}_{i,j}$ is placed on the $(\ell,\mathcal{H}_{\ell,j})$-th cell of the array.

\item By construction, there are precisely ${n_h-1 \choose \nu+s-1}$ layers at which the array is filled up for every column $j$. We obtain the $(b \times n_h)$ codeword array $\underline{c}_i$ by omitting unfilled cells of the $(L \times n_h)$ array generated in the previous step, where $b = {n_h-1 \choose \nu+s-1}$.
\een 
This completes the description of encoding $\underline{g}_i$ into $\underline{c}_i$ and clearly it is a 1-1 correspondence. Since the codeword array involves a layering of MDS codewords, the code is referred to as a {\em layered code} with parameter $\nu$. The edge node can complete the encoding process to generate $\underline{c}_i$ and transmit the $j$-th column $\underline{c}_{i,j} \in \mathbb{A}^b$ to the helper $H_j$. 
Alternatively, the edge node $E_i$ can start transmission even before computing the entire codeword array. This is possible because the coded symbols are generated layer-by-layer, and once encoding within a layer is finished, those symbols can be transmitted to respective helpers. The algorithm followed by each edge node to transmit the codeword array in this manner is given in Alg.~\ref{alg:layered_scheme}.

\begingroup
%\removelatexerror
\begin{algorithm}%[H]
	\DontPrintSemicolon
	
	Let $\mathbf{G}\in \mathbb{F}^{\nu\times (\nu+s)}$ denote a generator matrix of a $[\nu+s,\nu]$-MDS code.
	
	\For{$\ell\in[{n_h \choose \nu+s}]$}    
	{ 
		\tcp{Encoding}
		Edge node $E_i$ generates $\nu+s$ coded fragments $\{\underline{c}^{(\ell)}_{i,j}\}_{j\in[\nu+s]}$ in the following manner:
		\beqn
		[\underline{c}^{(\ell)}_{i,1}\ \ \underline{c}^{(\ell)}_{i,2}\ \ \cdots \ \ \underline{c}^{(\ell)}_{i,\nu+s}]=[\underline{g}^{(\ell)}_{i,1}\ \ \underline{g}^{(\ell)}_{i,2}\ \ \cdots \ \ \underline{g}^{(\ell)}_{i,\nu}]\mathbf{G}.
		\eeqn
		
		\tcp{Transmission}
		Edge node-$i$ transmits the coded fragment $\underline{c}^{(\ell)}_{i,j}$ to helper node $\mathcal{H}_{\ell,j}$ $\forall j\in[\nu+s]$.
	}
	\caption{Algorithm used by edge node $E_i$ to transmit coded gradient symbols \label{alg:layered_scheme}}
\end{algorithm}
\endgroup

\subsubsection{An Example\label{sec:example}}

\begin{figure*}
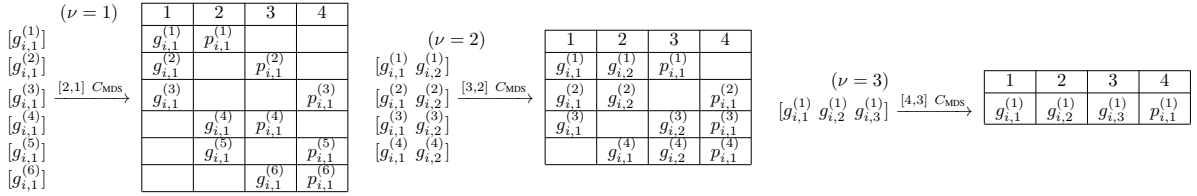

	\begin{center}
		\scalebox{0.7}{
			\begin{tabular}{l|c|c|c|c|}
				\cline{2-5} 
				$\hspace{1cm}(\nu=1)$& $1$ & $2$ & $3$ & $4$  \\
				\cline{2-5} 
				$[g_{i,1}^{(1)}]$  & $g_{i,1}^{(1)}$ & $p_{i,1}^{(1)}$ &  &  \\
				\cline{2-5} 
				$[g_{i,1}^{(2)}]$  & $g_{i,1}^{(2)}$ & &  $p_{i,1}^{(2)}$   &  \\
				\cline{2-5} 
				$[g_{i,1}^{(3)}]$ $\xrightarrow[]{[2,1]{ \ C_\text{MDS}}}$ & $g_{i,1}^{(3)}$ & & &  $p_{i,1}^{(3)}$   \\
				\cline{2-5}
				$[g_{i,1}^{(4)}]$  & & $g_{i,1}^{(4)}$ & $p_{i,1}^{(4)}$ &   \\
				\cline{2-5} 	
				$[g_{i,1}^{(5)}]$  & & $g_{i,1}^{(5)}$ & & $p_{i,1}^{(5)}$   \\
				\cline{2-5}
				$[g_{i,1}^{(6)}]$  & & & $g_{i,1}^{(6)}$ & $p_{i,1}^{(6)}$ \\
				\cline{2-5}
		\end{tabular}}
		\scalebox{0.7}{
			\begin{tabular}{l|c|c|c|c|} 
				\cline{2-5} 
				$\hspace{1cm}(\nu=2)$ & $1$ & $2$ & $3$ & $4$  \\
				\cline{2-5} 
				$[g_{i,1}^{(1)} \ g_{i,2}^{(1)}]$  & $g_{i,1}^{(1)}$ & $g_{i,2}^{(1)}$ & $p_{i,1}^{(1)}$ &  \\
				\cline{2-5} 
				$[g_{i,1}^{(2)} \ g_{i,2}^{(2)}]$ $\xrightarrow[]{[3,2]{ \ C_\text{MDS}}}$ & $g_{i,1}^{(2)}$ & $g_{i,2}^{(2)}$  & & $p_{i,1}^{(2)}$ \\
				\cline{2-5} 
				$[g_{i,1}^{(3)} \ g_{i,2}^{(3)}]$ & $g_{i,1}^{(3)}$ & & $g_{i,2}^{(3)}$ &  $p_{i,1}^{(3)}$  \\
				\cline{2-5}
				$[g_{i,1}^{(4)} \ g_{i,2}^{(4)}]$  & & $g_{i,1}^{(4)}$ & $g_{i,2}^{(4)}$ & $p_{i,1}^{(4)}$  \\
				\cline{2-5} 	
		\end{tabular}}
		\scalebox{0.7}{
			\begin{tabular}{l|c|c|c|c|} 
				\cline{2-5} 
				$\hspace{1cm}(\nu=3)$ & $1$ & $2$ & $3$ & $4$  \\
				\cline{2-5} 
				$[g_{i,1}^{(1)} \ g_{i,2}^{(1)} \ g_{i,3}^{(1)}]$ $\xrightarrow[]{[4,3]{ \ C_\text{MDS}}}$ & $g_{i,1}^{(1)}$ & $g_{i,2}^{(1)}$ & $g_{i,3}^{(1)}$ & $p_{i,1}^{(1)}$  \\
				\cline{2-5}
		\end{tabular}}	
		\caption{The encoding at edge node $E_i$ as the parameter $\nu$ varies. In each matrix, a column represents the helper node to which transmission happens, and the row represents the layer. Observe that $b=3$ for $\nu\in[2]$ and $b=1$ for $\nu=3$. Here $n_h=4$ and $s=1$. Thus, $\nu \in [n_h-s]\triangleq [3]$.\label{fig:example}}  
	\end{center}
\end{figure*}

An example for the encoding process up to the second step is given in Fig.~\ref{fig:example} for system parameters $(n_h=4,s=1)$. The code parameter $\nu$ can take values in $\{1,2,3\}$, and the process is explained for each $\nu$. For each $\nu$, a systematic code $\mathcal{C}_\nu$ is considered by the edge nodes. The symbols $p_{i,1}^{(\ell)}, \ell \in [L]$ are the parity symbols generated by the code. The third step is omitted as it just involves stacking all non-empty cells of the array. 

\subsection{Description of ${\cal A}_\nu$\label{sec:agg}}

The aggregation strategy is tailored for the proposed layered code ${\cal C}_{\nu}$ and aggregation is performed in each layer.

Suppose an $(n_e \times n_h)$ erasure matrix $\epsilon \in \Omega(s)$ has occurred, and $\epsilon(i,j)$ denotes the $(i,j)$-th entry of $\epsilon$. Let ${\cal E}_i = \{j \mid \epsilon(i,j)=1 \} \subseteq [n_h]$ denote the subset of helpers that failed to receive transmissions from $E_i$. Recall that $\underline{c}_{i} \in \mathbb{A}^{b\times n_h}$ can be viewed as an $(L \times n_h)$ array in which each row is called a layer. The $(\ell,j)$-th cell contains one of the $\underline{c}_{i,l}^{(\ell)}$'s if $j \in {\cal H}_{\ell}$, or else is empty. 

At every helper $H_j$, aggregation involves computing partial sums of coded symbols belonging to the same layer, but coming from different edges. Let us fix $\ell \in [L]$ with an associated subset ${\cal H}_{\ell} \subseteq [n_h]$. In what follows, we first develop some notation and subsequently explain the aggregation for layer $\ell$. Set $\alpha = {\nu+s \choose s}$, and let ${\cal S}^{(\ell)} = \{{\cal S}^{(\ell)}_1, {\cal S}^{(\ell)}_2, \ldots, {\cal S}^{(\ell)}_{\alpha}\}$ be the collection of $s$-element subsets of ${\cal H}_{\ell}$. We assume a lexicographic ordering on the set ${\cal S}^{(\ell)}$.  Next, we define a relation among the set of edges as follows; an edge $E_a$ is related to $E_b$ if ${\cal E}_a \cap {\cal H}_{\ell} = {\cal E}_b \cap {\cal H}_{\ell}$. By slight abuse of notation, we shall as well say $a$ is related to $b$ in the same manner. Clearly, it is an equivalence relation, and it splits $[n_e]$ into a partition ${\cal I}^{(\ell)} = \{{\cal I}^{(\ell)}_1, {\cal I}^{(\ell)}_2,\ldots, {\cal I}^{(\ell)}_{\Delta^{(\ell)}}\}$ where it is clear that $\Delta^{(\ell)} \leq n_e$.  Finally we define a mapping $\phi:{\cal I}^{(\ell)} \rightarrow {\cal S}^{(\ell)}$ as:
\bean 
\phi({\cal I}^{(\ell)}_\delta) & = & \min \Psi({\cal I}^{(\ell)}_\delta), 
\eean 
where $\Psi({\cal I}^{(\ell)}_\delta) = \{ {\cal S}^{(\ell)}_r \mid {\cal H}_{\ell} \cap {\cal E}_i \subseteq {\cal S}^{(\ell)}_r \text{ for some } i \in {\cal I}^{(\ell)}_\delta \}$ is a subset of ${\cal S}^{(\ell)}$. Since ${\cal E}_a \cap {\cal H}_{\ell} \subseteq {\cal H}_{\ell}$ and $|{\cal E}_a \cap {\cal H}_{\ell}| \leq |{\cal E}_a| \leq s$ for any $a \in [n_e]$, the set $\Psi({\cal I}^{(\ell)}_\delta)$ is always non-empty. The minimum is thus well-defined by the lexicographic ordering. Hence, $\phi$ is well-defined as well. Let us use $\phi({\cal I}^{(\ell)})$ to denote the image of $\phi$ in ${\cal S}^{(\ell)}$ and $\phi^{-1}({\cal S}^{(\ell)}_a)$ to denote the pre-image of any ${\cal S}^{(\ell)}_a \in \phi({\cal I}^{(\ell)})$. Let us define:
\bean
\beta^{(\ell)} = |\phi({\cal I}^{(\ell)})|
\eean
as the size of the image of $\phi$. Since $|{\cal I}^{(\ell)}| \leq n_e$ and $|{\cal S}^{(\ell)}| = \alpha$, clearly $\beta^{(\ell)} \leq \min\{n_e,\alpha \}$. Without loss of generality we enumerate $\phi({\cal I}^{(\ell)})$ as ${\cal S}^{(\ell)}_1, {\cal S}^{(\ell)}_2, \ldots, {\cal S}^{(\ell)}_{\beta^{(\ell)}}$.
 
We note that ${\cal I}^{(\ell)}_\delta$ is a subset of edges whereas ${\cal S}^{(\ell)}_a=\phi({\cal I}^{(\ell)}_\delta)$ is a subset of helpers. Thus the mapping $\phi$ associates a set of edges to a set of helpers, and thereby plays an important role in identifying which helpers should aggregate symbols from which edges. With all notations in place, we provide the aggregation algorithm ${\cal A}_{\nu,j}$ employed by $H_j$ as a pseudo-code in Alg.~\ref{alg:agg}. The algorithm generates $\underline{v}_j$ to be transmitted as its output, and the value of variable $m$ at the end of execution is precisely $m_j(\epsilon)$. 

%$\underline{c}_{i,j} \in \mathbb{A}^b$ is transmitted from $E_i$ to $H_j$. The single-max aggregation ${\cal A}_{\text{single}}$ is determined by ${\cal A}_j, j=1,2,\ldots, n_h$, and in turn by how transmitted vector $\underline{v}_j = [v_{j,1}, \ v_{j,2}, \ldots, v_{j,m_j(\epsilon)}] \in \mathbb{A}^{m_j(\epsilon)}$ from $H_j$ is determined. The aggregation algorithm used by $H_j$ to generate $\underline{v}_j$ is given in Alg.~\ref{alg:agg}. The multi-max strategy ${\cal A}_{\mu\text{-max}}$ is parameterized by $\mu \in \{1,2,\ldots, \alpha\}$ and it differs from ${\cal A}_{\text{single}}$ only at \textsl{Lines} $5$-$7$ in Alg.~\ref{alg:agg}. In contrast to considering $\ell=1$ alone, we shall compute $\sum_{i \in {\cal R}_{\ell}} \underline{c}_{i,j}$ for $\ell=1,2,\ldots, \mu$, provided the symbols are unerased by the channel.

\begingroup
%\removelatexerror
\begin{algorithm}%[H]
	\DontPrintSemicolon
	
	Erasure matrix $\epsilon$ is observed. Initialize $\underline{v}_j$ as empty vector, $m=0$.
	
	\For{$\ell \in 1, 2,\ldots L$} 
	{ 
		Compute ${\cal S}^{(\ell)}$, ${\cal I}^{(\ell)}$ and $\phi({\cal I}^{(\ell)})$
		
		Set $\beta^{(\ell)} = |\phi({\cal I}^{(\ell)})|$
		
		\For{$a \in 1,2,\ldots, \beta^{(\ell)}$}{
			
			\If{$j \in {\cal H}_{\ell} \setminus S^{(\ell)}_a$}{
				Compute ${\cal I} = \cup_{I \in \phi^{-1}(S^{(\ell)}_a)} I$
				
				$m \leftarrow m+1$
					
			    \tcp{aggregate symbols from ${\cal I}$}				
				Compute $v_{j,m} = \sum_{i \in {\cal I}} \underline{c}^{(\ell)}_{i,j}$ 
				
				\tcp{append $v_{j,m}$ to $\underline{v}_j$}
				$\underline{v}_j  \leftarrow \left[ \underline{v}_j ; v_{j,m}\right] $
			}
		}
	}
	\caption{Strategy ${\cal A}_{\nu,j}$ adopted by $H_j$ to generate aggregated symbols \label{alg:agg}}
\end{algorithm}
\endgroup

\subsubsection{An Example} Let $(n_e=7,n_h=6,s=2)$ and the code has parameter $\nu=2$. Consider layer $\ell$ that is described in \eqref{eq:al1}.  
\bea \label{eq:al1}
	\begin{array}{|c|c|c|c|c|c|} 
	\hline  
	1 & 2 & 3 & 4 & 5 & 6  \\
	\hline 
	g_{i,1}^{(\ell)} & g_{i,2}^{(\ell)} &  p_{i,1}^{(\ell)} &  p_{i,2}^{(\ell)} & & \\
	\hline 
\end{array}
\eea
Observe that ${\cal H}_{\ell} = \{1,2,3,4\}$. Let the erasure matrix $\epsilon$ be as given in \eqref{eq:al2}.
\bea \label{eq:al2}
\begin{array}{c|c|c|c|c|c|c|} 
	\cline{2-7}  
	& 1 & 2 & 3 & 4 & 5 & 6  \\
	\hline 
	E_1 & 0 & 0 & 0 & 0 & 1 & 1 \\
	\cline{2-7} 
	E_2 & 0 & 0 & 0 & 0 & 1 & 1 \\
\cline{2-7}
	E_3 & 0 & 0 & 0 & 1 & 1 & 0 \\
\cline{2-7}
	E_4 & 0 & 0 & 1 & 1 & 0 & 0 \\
\cline{2-7}
	E_5 & 0 & 0 & 1 & 1 & 0 & 0 \\
\cline{2-7}
	E_6 & 1 & 1 & 0 & 0 & 0 & 0 \\
\cline{2-7}
	E_7 & 1 & 1 & 0 & 0 & 0 & 0 \\
\cline{2-7}
\end{array}
\eea
Then the partition of edges ${\cal I}^{(\ell)}$ consists of  ${\cal I}^{(\ell)}_1 = \{1,2\}$, ${\cal I}^{(\ell)}_2 = \{3\}$, ${\cal I}^{(\ell)}_3 = \{4,5\}$ and ${\cal I}^{(\ell)}_4 = \{6,7\}$. They are mapped to ${\cal S}^{(\ell)}$, the set of 2-element subsets of  ${\cal H}_{\ell}$ as $\phi({\cal I}^{(\ell)}_1) = \{1,2\} := {\cal S}^{(\ell)}_1$, $\phi({\cal I}^{(\ell)}_2) = \{1,4\}:= {\cal S}^{(\ell)}_2$, $\phi({\cal I}^{(\ell)}_3) = \{3,4\}:= {\cal S}^{(\ell)}_3$ and $\phi({\cal I}^{(\ell)}_4) = \{1,2\}= {\cal S}^{(\ell)}_1$. We have $\beta^{(\ell)} = 3$. The aggregation at $H_1$ is done by executing ${\cal A}_{\nu,1}$. It can be seen that \textsl{Line 6} is \textsc{true} only while picking ${\cal S}^{(\ell)}_3 = \{3,4\}$, and $\phi^{-1}({\cal S}^{(\ell)}_3) = \{\{4,5\}\}$. Therefore, the current layer $\ell$ appends $c_{4,1}^{(\ell)} + c_{5,1}^{(\ell)}$ to $\underline{v}_1$.
 
\section{Worst-Case Communication Costs}

\subsection{Calculation of $C_\text{EH}$}

%For fixed $i\in[n_e]$ and $j\in[n_h]$, there are precisely ${n_h-1 \choose \nu+s-1}$ layers in which the edge node $i$ makes transmissions to the helper node $j$. As each tranmission is worth $\frac{p}{\lambda}$ symbols, the total communication bandwidth between an edge-helper-node pair is ${n_h-1 \choose \nu+s-1}\frac{p}{\lambda}$. Thus, the total communication from an edge node to {\it all} the helper nodes is given by (after normalizing with respect to $p$):
%\bean
%C_{EH} & = & n_h{n_h-1 \choose \nu+s-1}\frac{1}{\lambda}\\
%& = & n_h {n_h-1 \choose \nu+s-1} \frac{1}{{n_h \choose \nu+s}\nu} \\
%& = & %n_h\frac{(n_h-1)!}{(n_h-\nu-s)!(\nu+s-1)!}\frac{1}{\frac{n_h!}{%(n_h-\nu-s)!(\nu+s)!}\nu}\\
%& = & \frac{\nu+s}{\nu}.
%\eean

It is straightforward to compute 
\bean
\nonumber C_\text{EH} & = & \frac{n_hdb}{p} \ = \ \frac{n_h}{p} \cdot \frac{p}{\nu {n_h \choose \nu+s}} \cdot {n_h-1 \choose \nu+s-1}  \\
& = & \frac{\nu+s}{\nu} .
\eean 

\subsection{Calculation of $C_\text{HM}$}

Let the scheme be chosen with parameter $\nu$, and an erasure matrix $\epsilon$ be given.  In Alg.~\ref{alg:agg}, at most one symbol is generated inside the inner \textsl{for} loop at \textsl{Line 5}. Furthermore, if we consider $\{{\cal A}_{\nu,j}\mid j\in [n_h]\}$, then the \textsl{if} condition at \textsl{Line 6} is \textsc{true} for exactly $\nu$ of the $\mathcal{A}_{\nu,j}$'s, when $\ell$ is kept as a constant. By counting the number of times \textsl{Line 8} is executed in all of $\{{\cal A}_{\nu,j}\mid j\in [n_h]\}$ in two different ways, we obtain:
\bea
\nonumber \sum_{j=1}^{n_h}{m_j(\epsilon)} & = & \sum_{\ell \in [L]} \nu \beta^{(\ell)} \\
\label{eq:chmscheme} \Rightarrow \ \ \ C_{\text{HM}}(\epsilon) & = & \frac{d\nu \sum_{\ell \in [L]} \beta^{(\ell)}}{p} \ = \ \frac{1}{L}\sum_{\ell \in [L]} \beta^{(\ell)},
\eea
since $d=p/L\nu$. It may be noted that $\beta^{(\ell)}$ depends on the erasure matrix $\epsilon$ though it is not made explicit in its notation. The next lemma estimates the worst-case quantity $C_\text{HM}$.
\blem \label{lem:chmbound} Let $(p,n_e,n_h,s)$ be given. For a layered coding scheme with parameter $\nu$,
\bea \label{eq:chmbound} 
C_{\text{HM}} & \leq & \min\left\{n_e,{\nu+s\choose s}\right\}.
\eea
\elem 
\bpf In Sec.~\ref{sec:agg}, it is established that $\beta^{(\ell)} \leq  \min\left\{n_e,{\nu+s\choose s}\right\}$ for any $\ell$, independent of the erasure matrix $\epsilon$. Substituting in \eqref{eq:chmscheme}, the bound follows.
\epf 
In the following, we will argue that the bound in \eqref{eq:chmbound} is indeed tight when there is a sufficiently large number of edges. Assume $n_e \geq {n_h \choose s}$. Consider the erasure matrix $\epsilon^*$ in which every possible erasure pattern occurs in at least one edge. Then for every layer $\ell$, there is at least an edge $i$ for which ${\cal H}_{\ell} \cap {\cal E}_i = {\cal E}_i = {\cal S}^{(\ell)}_r$ for every $r\in[\alpha]$. This implies that $\phi$ is an onto function and as a result, $\beta^{(\ell)} = \alpha = {\nu+s \choose s}$ for the erasure matrix $\epsilon^*$. Substituting this in \eqref{eq:chmscheme} and by Lemma~\ref{lem:chmbound}, we have  the following theorem. 

\bthm Let $(p,n_e,n_h,s)$ be given. Let $\nu \in \{ 1,2,\ldots,n_h-s\}$. If $n_e \geq {n_h \choose s}$, then for a layered coding scheme with parameter $\nu$,
\bea \label{eq:chmeq} 
C_{\text{HM}} & = & {\nu+s\choose s}.
\eea
\ethm

For $\nu=1$, our client code is equivalent to that of the ARC scheme proposed in \cite{PrRPA2020} and achieves $C_\text{EH} = s+1$.  On the other end, when $\nu=n_h-s$, it reduces to that of the AMC scheme and achieves $C_\text{EH} = \frac{n_h}{n_h-s}$. In addition, when the parameter $\nu$ is varied over $[n_h-s]$, we achieve a graceful transition from the ARC scheme to the AMC scheme. For instance, we illustrate this transition in Fig.~\ref{fig:tradeoff} when $(n_e=50,n_h=10,s=2)$.

\bibliographystyle{IEEEtran}
\bibliography{cga}

\begin{thebibliography}{10}
\providecommand{\url}[1]{#1}
\csname url@samestyle\endcsname
\providecommand{\newblock}{\relax}
\providecommand{\bibinfo}[2]{#2}
\providecommand{\BIBentrySTDinterwordspacing}{\spaceskip=0pt\relax}
\providecommand{\BIBentryALTinterwordstretchfactor}{4}
\providecommand{\BIBentryALTinterwordspacing}{\spaceskip=\fontdimen2\font plus
\BIBentryALTinterwordstretchfactor\fontdimen3\font minus
  \fontdimen4\font\relax}
\providecommand{\BIBforeignlanguage}[2]{{%
\expandafter\ifx\csname l@#1\endcsname\relax
\typeout{** WARNING: IEEEtran.bst: No hyphenation pattern has been}%
\typeout{** loaded for the language `#1'. Using the pattern for}%
\typeout{** the default language instead.}%
\else
\language=\csname l@#1\endcsname
\fi
#2}}
\providecommand{\BIBdecl}{\relax}
\BIBdecl

\bibitem{McMRHA2017c}
H.~B. McMahan, E.~Moore, D.~Ramage, S.~Hampson, and B.~A. y~Arcas,
  ``{Communication-Efficient Learning of Deep Networks from Decentralized
  Data},'' in \emph{Proc. International Conference on Artificial Intelligence
  and Statistics (AISTATS)}, vol. 108, 2017, pp. 1273--1282.

\bibitem{LiSTS20}
T.~{Li}, A.~K. {Sahu}, A.~{Talwalkar}, and V.~{Smith}, ``{Federated Learning:
  Challenges, Methods, and Future Directions},'' \emph{IEEE Signal Processing
  Magazine}, vol.~37, no.~3, pp. 50--60, 2020.

\bibitem{ReMHP2020}
A.~Reisizadeh, A.~Mokhtari, H.~Hassani, A.~Jadbabaie, and R.~Pedarsani,
  ``{FedPAQ: A Communication-Efficient Federated Learning Method with Periodic
  Averaging and Quantization},'' in \emph{Proc. International Conference on
  Artificial Intelligence and Statistics (AISTATS)}, vol. 108, 2020, pp.
  2021--2031.

\bibitem{ZhWZZC2020}
L.~{Zhao}, Q.~{Wang}, Q.~{Zou}, Y.~{Zhang}, and Y.~{Chen},
  ``{Privacy-Preserving Collaborative Deep Learning With Unreliable
  Participants},'' \emph{IEEE Transactions on Information Forensics and
  Security}, vol.~15, pp. 1486--1500, 2020.

\bibitem{TaLDK2017}
R.~Tandon, Q.~Lei, A.~G. Dimakis, and N.~Karampatziakis, ``{Gradient Coding:
  Avoiding Stragglers in Distributed Learning},'' in \emph{Proc. International
  Conference on Machine Learning (ICML)}, vol.~70, 2017, pp. 3368--3376.

\bibitem{YeA2018}
M.~Ye and E.~Abbe, ``{Communication-Computation Efficient Gradient Coding},''
  in \emph{Proc. International Conference on Machine Learning (ICML)}, vol.~80,
  2018, pp. 5610--5619.

\bibitem{LiMMYSA2018}
S.~Li, S.~M.~M. Kalan, Q.~Yu, M.~Soltanolkotabi, and A.~S. Avestimehr,
  ``{Polynomially Coded Regression: Optimal Straggler Mitigation via Data
  Encoding},'' \emph{CoRR}, vol. abs/1805.09934, 2018.

\bibitem{RePPA2019}
A.~Reisizadeh, S.~Prakash, R.~Pedarsani, and A.~S. Avestimehr, ``{Tree Gradient
  Coding},'' in \emph{Proc. {IEEE} International Symposium on Information
  Theory (ISIT)}, 2019, pp. 2808--2812.

\bibitem{KHKJSIT21}
M.~N. Krishnan, E.~Hosseini, and A.~Khisti, ``{Sequential Gradient Coding for
  Packet-Loss Networks},'' \emph{IEEE Journal on Selected Areas in Information
  Theory}, vol.~2, no.~3, pp. 919--930, 2021.

\bibitem{PrRPA2020}
S.~{Prakash}, A.~{Reisizadeh}, R.~{Pedarsani}, and A.~S. {Avestimehr},
  ``{Hierarchical Coded Gradient Aggregation for Learning at the Edge},'' in
  \emph{Proc. IEEE International Symposium on Information Theory (ISIT)}, 2020,
  pp. 2616--2621.

\bibitem{LZSL20}
L.~Liu, J.~Zhang, S.~Song, and K.~B. Letaief, ``{Client-Edge-Cloud Hierarchical
  Federated Learning},'' in \emph{Proc. IEEE International Conference on
  Communications (ICC)}, 2020, pp. 1--6.

\bibitem{BTJSACCGA}
B.~Sasidharan and A.~Thomas, ``{Coded Gradient Aggregation: A Tradeoff Between
  Communication Costs at Edge Nodes and at Helper Nodes},'' \emph{IEEE Journal
  on Selected Areas in Communications}, vol.~40, no.~3, pp. 761--772, 2022.

\bibitem{HuaCL2007}
C.~Huang, M.~Chen, and J.~Li, ``{Pyramid Codes: Flexible Schemes to Trade Space
  for Access Efficiency in Reliable Data Storage Systems},'' in \emph{Proc.
  IEEE International Symposium on Network Computing and Applications (NCA)},
  2007, pp. 79--86.

\bibitem{BlaBV1996}
M.~Blaum, J.~Bruck, and A.~Vardy, ``{MDS array codes with independent parity
  symbols},'' \emph{IEEE Transactions on Information Theory}, vol.~42, no.~2,
  pp. 529--542, 1996.

\bibitem{DimGWWR2010}
A.~G. Dimakis, P.~B. Godfrey, Y.~Wu, M.~J. Wainwright, and K.~Ramchandran,
  ``Network coding for distributed storage systems,'' \emph{IEEE Transactions
  on Information Theory}, vol.~56, no.~9, pp. 4539--4551, 2010.

\bibitem{AlGLTV2017}
D.~Alistarh, D.~Grubic, J.~Li, R.~Tomioka, and M.~Vojnovic, ``{QSGD:
  Communication-Efficient SGD via Gradient Quantization and Encoding},'' in
  \emph{Proc. Advances in Neural Information Processing Systems (NeurIPS)},
  vol.~30, 2017, pp. 1709--1720.

\bibitem{AcDFS2019}
J.~Acharya, C.~De~Sa, D.~Foster, and K.~Sridharan, ``{Distributed Learning with
  Sublinear Communication},'' in \emph{Proc. International Conference on
  Machine Learning (ICML)}, vol.~97, 2019, pp. 40--50.

\bibitem{ShCEPC2020}
N.~Shlezinger, M.~Chen, Y.~C. Eldar, H.~V. Poor, and S.~Cui, ``{Federated
  Learning with Quantization Constraints},'' in \emph{Proc. {IEEE}
  International Conference on Acoustics, Speech and Signal Processing
  (ICASSP)}, 2020, pp. 8851--8855.

\bibitem{TBAPTIT}
C.~Tian, B.~Sasidharan, V.~Aggarwal, V.~A. Vaishampayan, and P.~V. Kumar,
  ``{Layered Exact-Repair Regenerating Codes via Embedded Error Correction and
  Block Designs},'' \emph{IEEE Transactions on Information Theory}, vol.~61,
  no.~4, pp. 1933--1947, 2015.

\bibitem{SPKVSK16}
B.~Sasidharan, N.~Prakash, M.~N. Krishnan, M.~Vajha, K.~Senthoor, and P.~V.
  Kumar, ``Outer bounds on the storage-repair bandwidth trade-off of
  exact-repair regenerating codes,'' \emph{International Journal of Information
  and Coding Theory}, vol.~3, no.~4, pp. 255--298, 2016.

\end{thebibliography}

\end{document}